\documentclass[sigconf,hdvipdfm,dvipsnames]{acmart}

\usepackage{booktabs,multirow} 

\usepackage{graphicx}
\usepackage{subfigure}

\usepackage{url}

\usepackage{balance}

\usepackage{xcolor,footnote}

\usepackage{stfloats}

\setcopyright{acmlicensed}

\copyrightyear{2017}
\acmYear{2017}
\acmConference{APSys '17}{September 2-3, 2017}{Mumbai, India}
\acmPrice{15.00}
\acmDOI{10.1145/3124680.3124730}
\acmISBN{978-1-4503-5197-3/17/09}

\begin{document}

\title{ACTS in Need: \underline{A}utomatic \underline{C}onfiguration \underline{T}uning\\with \underline{S}calability Guarantees}
\titlenote{Yuqing Zhu is the corresponding author.}

\author{Yuqing Zhu, Jianxun Liu, Mengying Guo, Wenlong Ma, Yungang Bao}
\affiliation{%
  \institution{Advanced Computer Systems Research Center}
  \streetaddress{Institute of Computing Technology, Chinese Academy of Sciences}
  \city{Beijing}
  \country{China}
}
\email{zhuyuqing,liujianxun,guomengying,mawenlong,baoyungang@ict.ac.cn}

\begin{abstract}
To support the variety of Big Data use cases, many Big Data related systems expose a large number of user-specifiable configuration parameters. Highlighted in our experiments, a MySQL deployment with well-tuned configuration parameters achieves a peak throughput as \textbf{12 times} much as one with the default setting. However, finding the best setting for the tens or hundreds of configuration parameters is mission impossible for ordinary users. Worse still, many Big Data applications require the support of multiple systems co-deployed in the same cluster. As these co-deployed systems can interact to affect the overall performance, they must be tuned together. Automatic configuration tuning with scalability guarantees (ACTS) is in need to help system users. Solutions to ACTS must scale to various systems, workloads, deployments, parameters and resource limits. Proposing and implementing an ACTS solution, we demonstrate that ACTS can benefit users not only in improving system performance and resource utilization, but also in saving costs and enabling fairer benchmarking.
\end{abstract}

\maketitle

\section{Introduction}

The Big Data industry is estimated to be worth more than hundreds of billions of dollars and still growing~\cite{bigdata,bigdata2}. Along with the Big Data phenomenon, many systems emerge to fulfill the tasks of collecting, processing and analyzing the huge amount of data, e.g., Hadoop~\cite{hadoop} and Spark~\cite{spark}. To support the variety of Big Data use cases, many Big Data related systems are designed and developed with a large number configuration parameters  (or knobs)~\cite{knobs}. For example, Hadoop~\cite{hadoop} has more than 180 knobs, while the database system MySQL~\cite{mysql} has more than 450 knobs. These tunable configuration parameters control nearly all aspects of system runtime behaviors~\cite{ituned}.

On the one hand, these configuration parameters are highly correlated with the system performance~\cite{tuneSpark,tuneHadoop,tuneMysql}. Take MySQL for instance. Changing the configuration setting can result in more than \textbf{11 times} performance gain for MySQL ($\S$\ref{sec:11times}). On the other hand, the large number of configuration parameters lead to an ever-increasing complexity of configuration issues that overwhelm users, developers and administrators. As multiple systems can be involved in a task of Big Data management, tuning multiple systems' configuration parameters that intrinsically interact in an application has surpassed the abilities of humans~\cite{asilomar}.

\begin{figure*}
  \centering
  \subfigure[MySQL: \textbf{uniform read}]{
    \label{fig:mysql} 
    \includegraphics[width=0.3\textwidth,height=90pt]{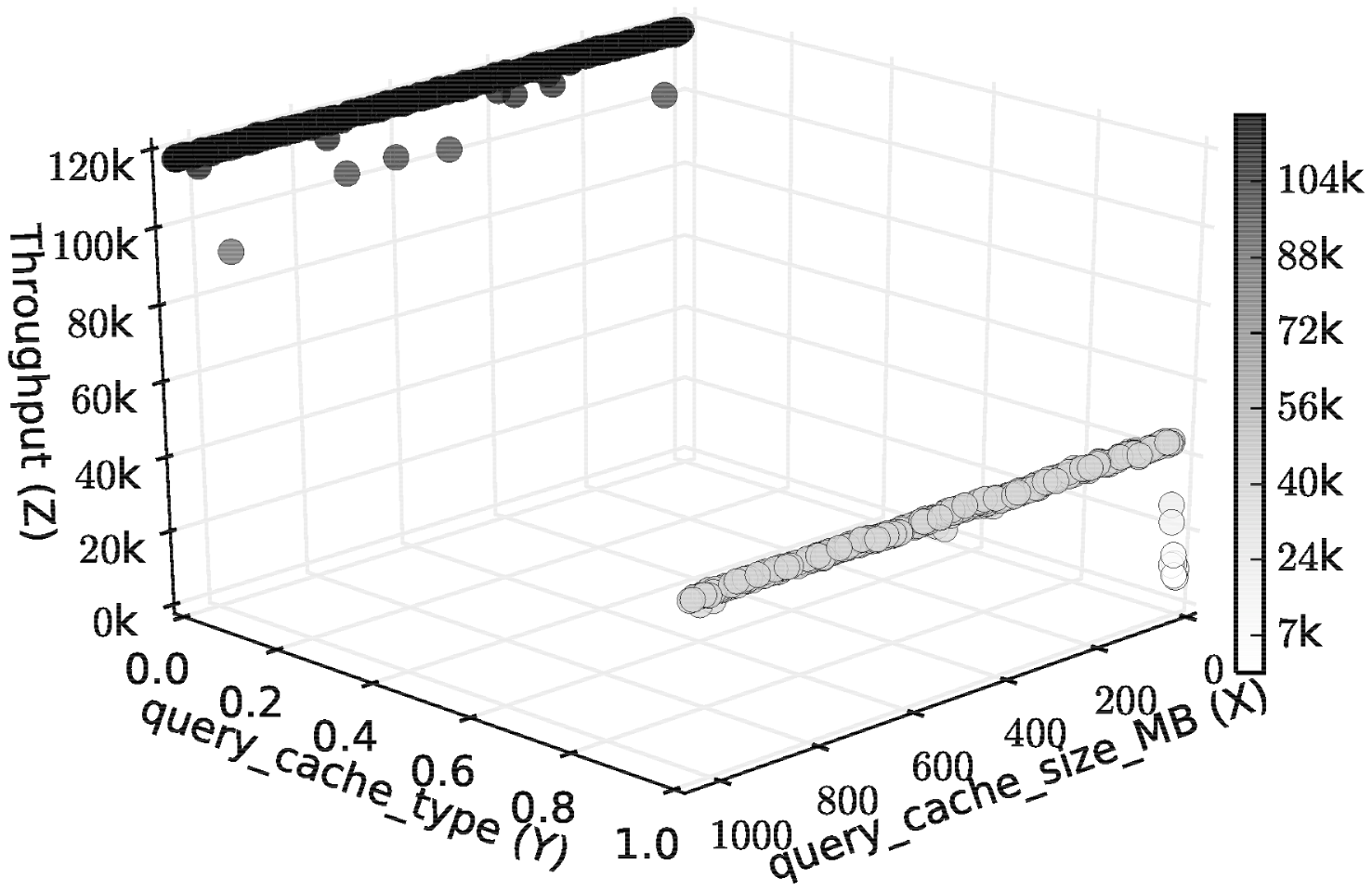}}
  \subfigure[Tomcat: \textbf{default JVM settings}]{
    \label{fig:tomcat} 
    \includegraphics[width=0.35\textwidth,height=90pt]{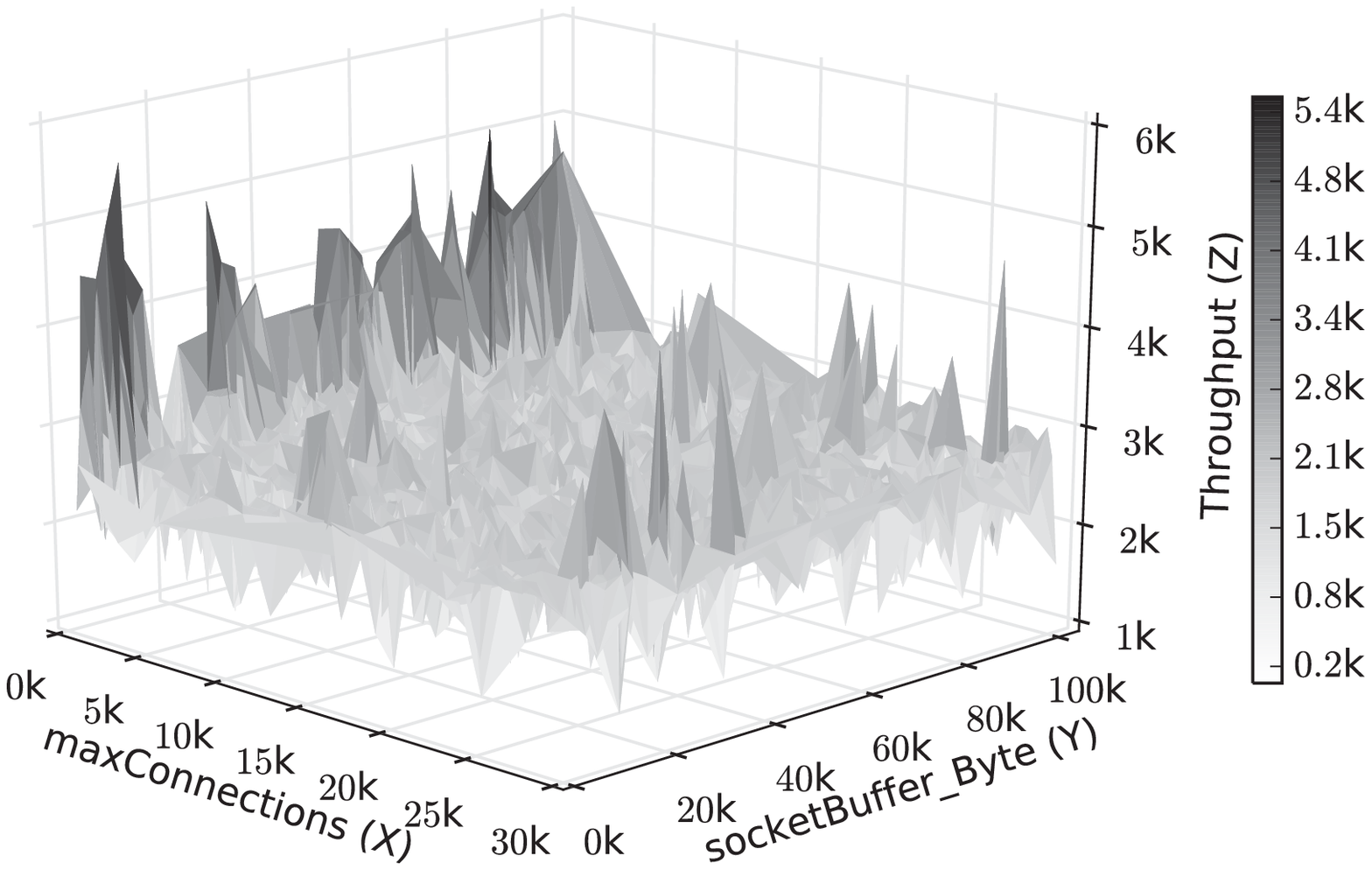}}
    \subfigure[Spark: \textbf{standalone}]{
    \label{fig:spark} 
    \includegraphics[width=0.28\textwidth,height=90pt]{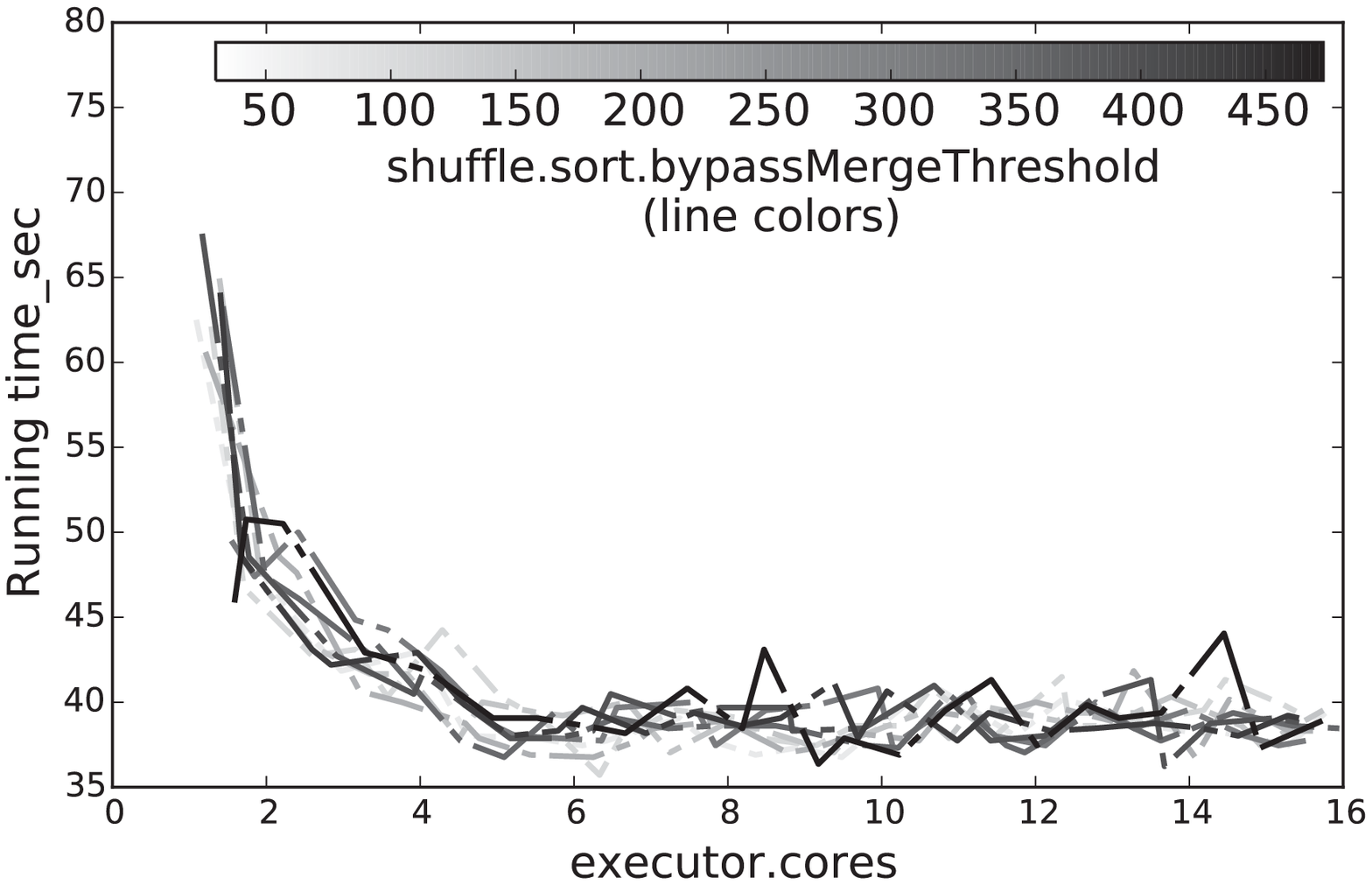}}
  \subfigure[MySQL: \textbf{zipfian read-write}]{
    \label{fig:mysqlZipf} 
    \includegraphics[width=0.31\textwidth,height=90pt]{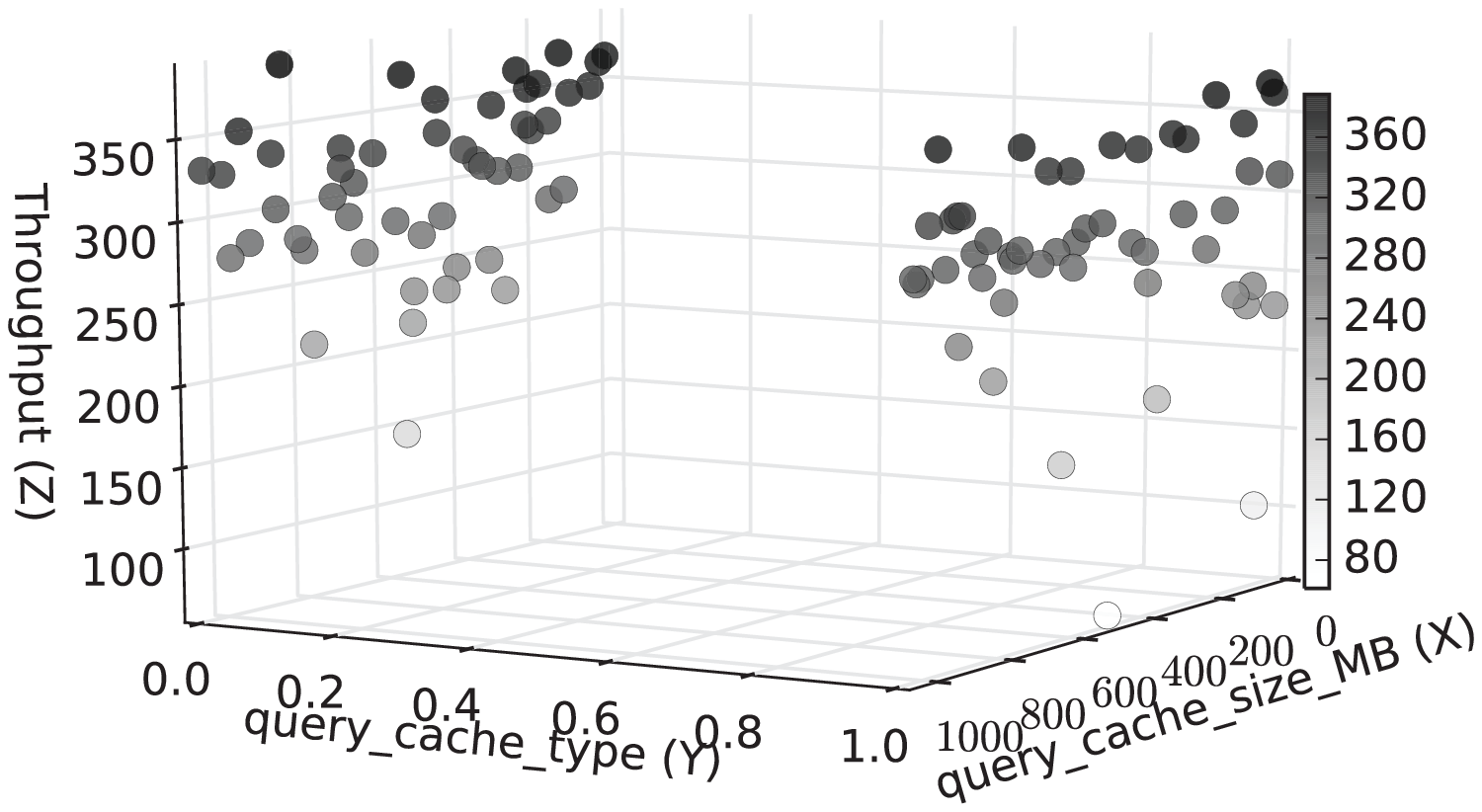}}
  \subfigure[Tomcat: \textbf{tuned JVM settings}]{
    \label{fig:tomcatJvm} 
    \includegraphics[width=0.35\textwidth,height=90pt]{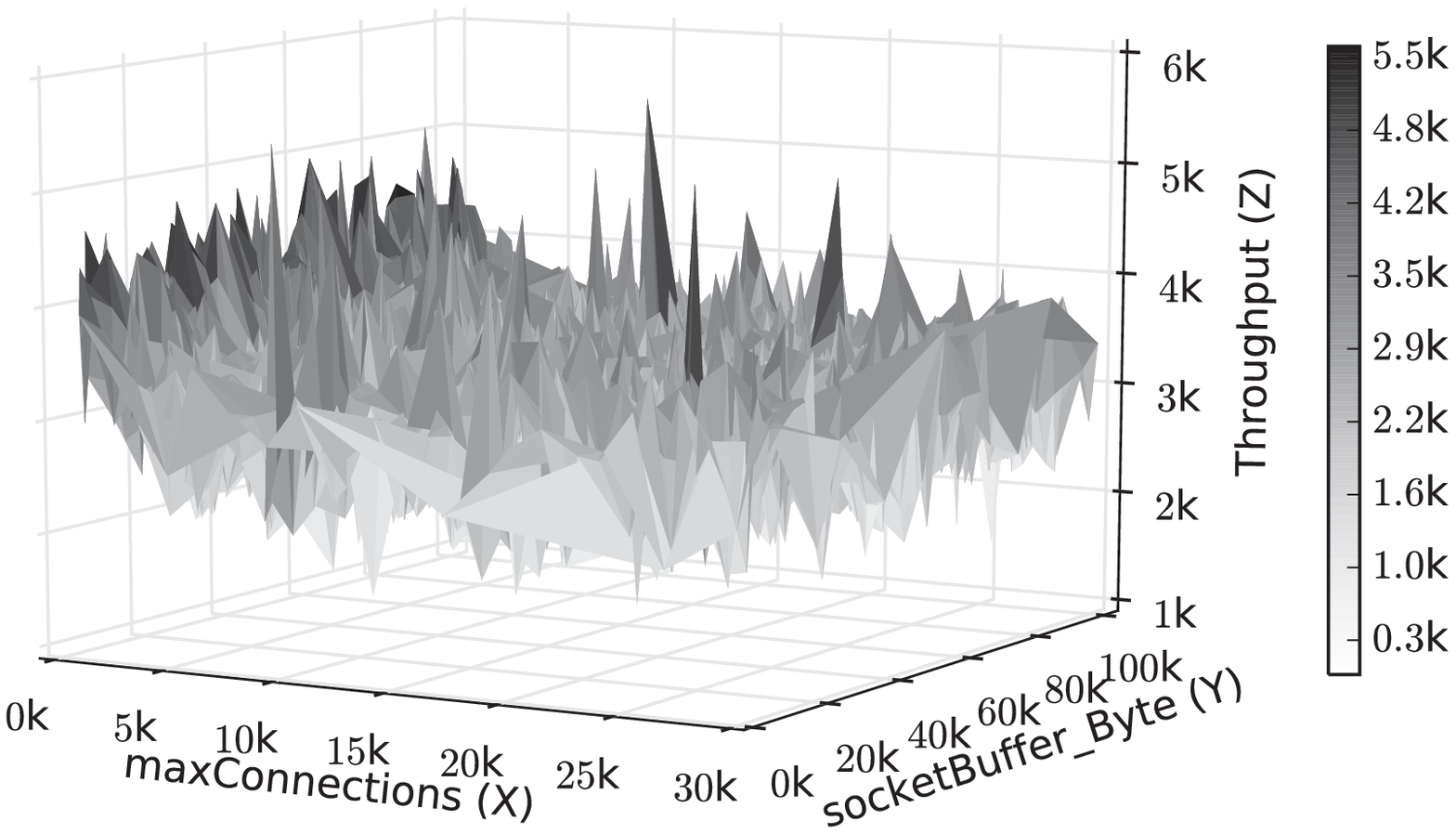}}
  \subfigure[Spark: \textbf{cluster}]{
    \label{fig:sparkcluster} 
    \includegraphics[width=0.29\textwidth,height=90pt]{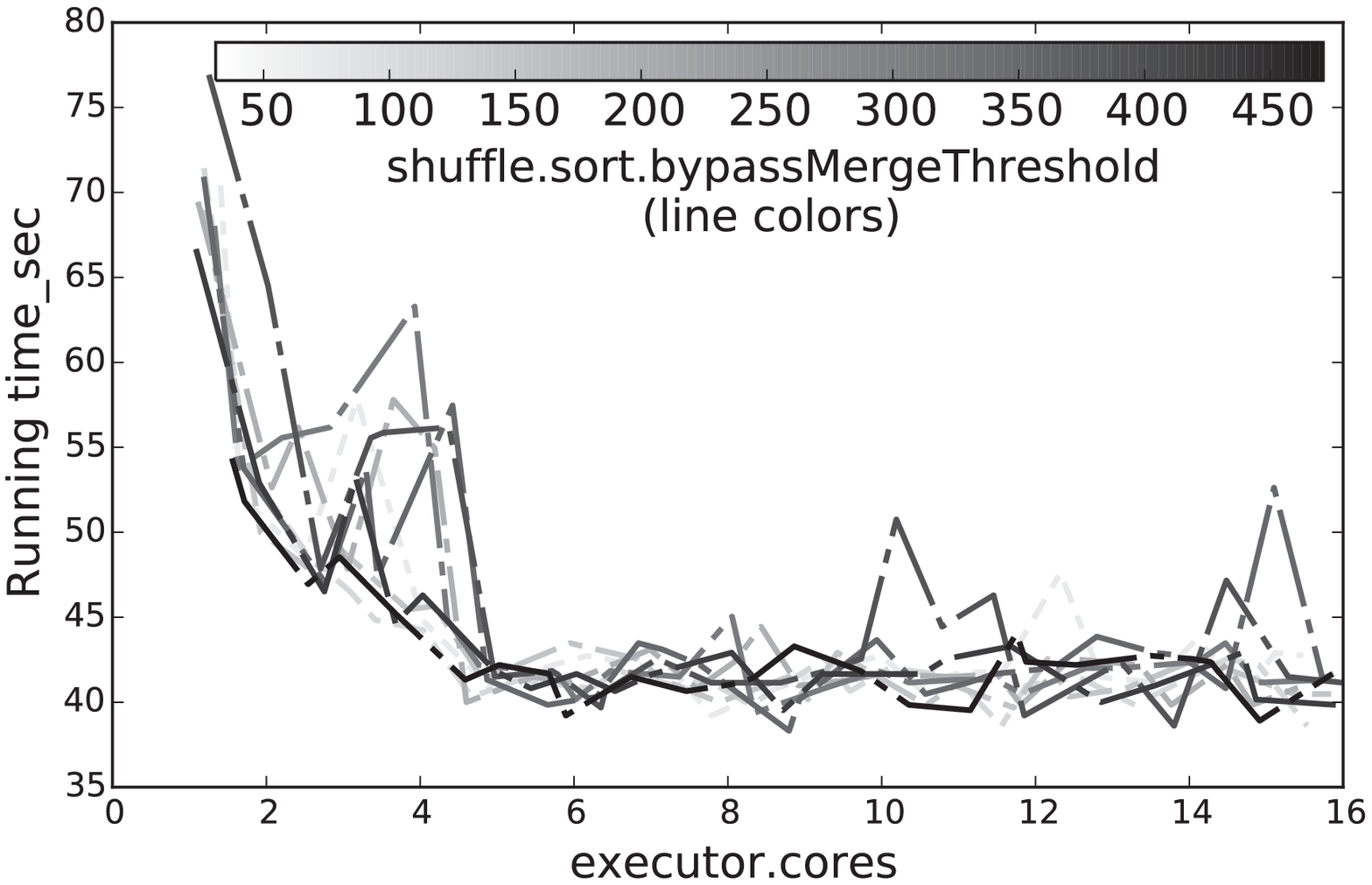}}
  \caption{\underline{Diverging} performance surfaces of \emph{MySQL}, \emph{Tomcat} and \emph{Spark} under different \underline{workloads} and \underline{deployments}.}
  \label{fig:perfFunc} 
\end{figure*}
Automatic configuration tuning (ACT) can help users tune the large number of configuration parameters towards a better overall performance. ACT involves solving the performance optimization problem in the high-dimensional space of configuration parameters. Previous attempts to automate configuration tuning have used search-based~\cite{rrs,smarthillclimbing}, model-based~\cite{starfish,ituned} or control-based~\cite{RLweb,virtualConfig} methods. Some of these methods assume manually constructed models~\cite{starfish} or the existence of system simulators~\cite{rrs,paraTuning,resSurf}. These assumptions are feasible for a specific system with a limited number of parameters~\cite{eurosysConf, RLweb}. Some methods assume the existence of a large sample set~\cite{aloja}. Many have not considered the influence of workloads on tuning~\cite{ottertune}, and hardly any related work considers the deployment environment as a factor that affects configuration tuning.

Now, the problem of automatic configuration tuning is facing three new challenges not studied before ($\S$\ref{sec:challenges}). These challenges come from the large number of configuration parameters, the dynamicity and complexity of the performance model, and the expensive sample collection process. The unprecedentedly large number of configuration parameters have complicated the performance model of a system. Worse still, the complex performance model is also related to factors like workloads and deployment environments; hence, a system must be tuned in a deployment environment similar or the same as the real system deployment. This fact makes the collection of samples very expensive.

Automatic configuration tuning with scalability guarantees (ACTS) is in need to address these challenges. In this work, we motivate the study of the ACTS problem: automatically tuning the system configuration parameters while addressing the above challenges by guaranteeing five scalability requirements ($\S$\ref{sec:problem}). The five scalability requirements involve the scalability regarding SUT (system under tune), workload, deployment environment, parameter set, and sample set size.

Despite the difficulty of the ACTS problem, we propose a preliminarily feasible solution ($\S$\ref{sec:solution}). This solution has a flexible system architecture different from previously proposed architectures. This flexible architecture can adapt to different SUTs, workloads and deployment environments. It also allows the easy integration of scalable sampling methods and scalable optimization algorithms to solve the ACTS problem. We have implemented the solution to demonstrate the benefits of ACTS.

Solving the ACTS problem can lead to great benefits for users ($\S$\ref{sec:benefits}), including but not limited to facilitating the system usage, improving the system performance, increasing the system utilization, saving labor costs, fairer benchmarking results, identifying system bottlenecks, etc. As only sporadic efforts are found to study the problem of automatic configuration tuning, our goal here is mainly to motivate the study of the ACTS problem, as well as demonstrating the feasibility and the great benefits of solving ACTS.

\section{New Challenges of ACT}%
\label{sec:challenges}%

The problem of automatic configuration tuning (ACT) is becoming more challenging. On the one hand, ACT involves an optimization problem which is previously known to be difficult to solve. On the other hand, new challenges are emerging, thanks to Big Data.

\subsection{A Large Number of Knobs}\vspace{3pt}

\textbf{The number of configuration parameters is now tens, hundreds or more}. Previously, the number of variables is multiple in the optimization problem of configuration tuning~\cite{rrs,paraTuning,resSurf}. As Big Data systems are targeting a wide range of use cases, they can provide tens or hundreds of configuration parameters~\cite{knobs} for users in various use cases and with various deployment environments. Moreover, co-deployed systems might need to be tuned together for one use case such that the number of parameters to tune can further increase. For example, the tuning guides for Hadoop suggest tuning the configuration parameters of both the Hadoop (with more than a hundred knobs) and the JVM (Java Virtual Machine, with tens of knobs)~\cite{hadoopJvm1,hadoopJvm2}. Thus, the performance optimization problem of configuration tuning must be solved in a high-dimensional space that rarely occurs in previous optimization problems' settings.\vspace{3pt}

\textbf{The number of parameters can hardly be reduced in configuration tuning}. The impacts of configuration parameters on a system's performance are intrinsic and complicated. Sometimes, a configuration parameter can have impacts unfound even by system developers~\cite{ituned}. Besides, configuration parameters are non-stochastic variables. Hence, the dimension reduction methods commonly used in machine learning cannot be applied to reduce the number of configuration parameters, e.g., factor analysis~\cite{fa} or principal component analysis~\cite{pca}. Even though the more impacting knobs can be tuned first~\cite{ottertune}, the less impacting knobs cannot be neglected in tuning, because it is likely that the combined impact of all the less impacting knobs exceeds that of the more impacting knobs.

\subsection{Dynamicity and Complexity}

We have carried out \textbf{thousands of experiments} to study the dynamicity and complexity of performance models. The performance model of an SUT is highly dynamic and complex, because it is related to not only the SUTs, but also the varying workloads and deployment environments. The impacts of the deployment environment can come from the hardware and the software~\cite{configInteracts1,configInteracts2}. Such impacts make it infeasible to decompose the SUT and the deployment environment into subcomponents for tuning. These facts also make it very difficult for human beings to manually construct models or simulators for general systems.\vspace{3pt}

\textbf{Different SUTs have different performance models}. Take the widely used database system MySQL~\cite{mysql}, Web server system Tomcat~\cite{tomcat} and big data processing system Spark~\cite{spark} for example. We plot in Figure~\ref{fig:mysql}, \ref{fig:tomcat} and \ref{fig:spark} their performance functions projected in low-dimensional spaces respectively. For MySQL, the projection is two lines, while Tomcat's is a irregularly bumpy surface. Spark's is a relatively smooth surface, which can be depicted as smooth lines when projected to a 2-D space.\vspace{3pt}

\textbf{Different workloads also lead to different performance models}. For the same deployment of MySQL, we apply the different workloads of uniform read and zipfian read-write. The differed workloads result in the diverging plots in Figure~\ref{fig:mysql} and \ref{fig:mysqlZipf}. For the uniform read workload, the \emph{query\_cache\_type} is the configuration parameter dominating the system performance. But for the zipfian read-write workload, the value of \emph{query\_cache\_type} has no such dominant influence. The impacts of configuration parameters are workload related.\vspace{3pt}

\textbf{The hardware of the deployment environment influences the performance model}. A system can be deployed on a single server or a server cluster. The system deployed on a single server generally behaves differently from that deployed in a cluster. To demonstrate the hardware impact from the deployment environment, we deploy Spark in the standalone mode and the cluster mode. Applying the same workload, we get two differed performance functions as plotted in Figure~\ref{fig:spark} and \ref{fig:sparkcluster}. As compared to the smooth performance function of the standalone mode, that of the cluster mode rises up sharply at some points, e.g., when the value of \emph{executor.cores} equals to four.\vspace{3pt}

\textbf{The co-deployed software also has intrinsic impacts on the SUT's performance}. For Big Data applications, multiple systems might need to be deployed together to accomplish one task. For example, we might need to deploy the Hadoop file system for using Spark; and, running Java-based systems requires the running of JVM (Java Virtual Machine). Co-deployed software systems can interact with and influence each other, as they might share hardware resources like CPU cycles, memory and network bandwidth. For instance, Figure~\ref{fig:tomcatJvm} differs from Figure~\ref{fig:tomcat} only in that we change the JVM setting \emph{TargetSurvivorRatio} when generating Figure~\ref{fig:tomcatJvm}. Although the performance surface remains as bumpy, the maximum performance is achieved at different areas.\vspace{-3pt}

\subsection{Costly Sample Collection}

\textbf{Only a limited number of samples can be collected for tuning}. Because of the impacts from the deployment environment and the workload, the performance-configuration samples can only be generated in tests that apply the workload on the deployed system. Thus, collecting a large set of tuning samples is too expensive to be practical. As performance models or simulators can hardly be constructed due to complexity, no arbitrary number of samples can be generated for configuration tuning. It is not practical at all to collect thousands of samples as required by existing solutions to configuration tuning~\cite{paraTuning}. Rather, users might expect a solution exploiting only hundreds or tens of samples, considering that Big Data workloads generally take time to run~\cite{cloudsuite,bigbench}. In sum, configuration tuning must restrain the overhead of sample collection.
\section{The New Problem: ACTS}\vspace{3pt}%
\label{sec:problem}%

The new challenges in fact call for solutions to a new problem. The new problem is the problem of \emph{automatic configuration tuning with scalability guarantees (ACTS)}. The ACTS problem is to find, within a given \textbf{resource limit}, a \textbf{configuration setting} that can optimize the performance of a given \textbf{SUT}'s \textbf{deployment} under a specific \textbf{workload}.\vspace{6pt}

Resource limit can be represented as the time or the number of tests allowed for tuning. Different measures for resource can be transformed into and represented by each other. For convenience, we consider in this paper the resource limit as the number of allowed tests, which is equal to the number of samples to be collected.

The solution to the ACTS problem must guarantee scalability with regard to \textbf{resource limit}, \textbf{configuration parameter set}, \textbf{SUT}, \textbf{deployment environment} and \textbf{workload}. When the resource limit is relaxed, the solution to ACTS is expected to output a configuration setting with a better performance. The solution must also be able to find new best configuration settings when new configuration parameter sets, SUTs, deployment environments and workloads are provided. It must adapt to the changes of these factors. Besides, the integration of evolved SUTs, deployment environments and workloads must be facilitated.\vspace{3pt}

The problem of ACTS and the scalability requirements invalidate the assumptions of related works on automatic configuration tuning. First, preconstructing models or simulators~\cite{starfish} has surpassed the capabilities of humans due to the large number of configuration parameters in the overall system, as well as due to the complicated interactions between the SUT, the workload and the deployment environment. Second, the sample collection becomes very expensive as the tuning samples can only be collected for a specific deployment of system, instead of being reused across deployments~\cite{ottertune}. The costs of sample collection must be taken into account in configuration tuning, rather than assuming a large sample set~\cite{paraTuning}. Third, assuming strong conditions for tuning is impractical as the performance model of an SUT is correlated with the varying workloads, hardware settings, co-deployed software, and the set of configuration parameters.

\section{A Preliminary ACTS Solution}%
\label{sec:solution}%

Although the ACTS problem is difficult, we demonstrate that it is solvable by presenting a preliminary ACTS solution in this section.

\subsection{Design Rationale}%
\label{sec:design}

To solve the ACTS problem, we must allow the tuning system to collect samples directly from the SUT in the target deployment environment and under the real workload. The sample collection process requires changing the configuration settings of the SUT, which must be restarted to allow the new configuration setting to take effect. Therefore, the tuning system must be able to control the SUT and run the workload. To fulfill this purpose, we design the architecture of the tuning system with the components of system manipulator and workload generator. To avoid the interference with the real applications on tuning, the design of this architecture takes advantage of the staging environment that commonly exists and that is the same as the actual deployment environment. The resulting architecture is plotted in Figure~\ref{fig:act4gen}. Section~\ref{sec:arch} presents a brief overview of the architecture and the interactions between system components.

While the problem about how to collect samples with scalability guarantees is solved by the flexible architecture, the problem about which samples to collect remains to be addressed. Due to the large number of configuration parameters and their wide ranges, it is impossible to try every possible combinations. In fact, only a very limited number of configuration settings can be tested and sampled, because of the resource limit in the ACTS problem. Here, there exists a subproblem of sampling.

The subproblem of sampling must handle all types of parameters, including boolean, enumeration and numerics. The resulted samples must have a wide coverage of the solution space. To guarantee scalability, the sampling method must also guarantee better coverage of the whole solution space if more samples are allowed by the users. Thus, the sampling method must produce sample sets satisfying the following three conditions: (1) the set has a wide coverage over the high-dimensional space of configuration parameters; (2) the set is small enough to meet the resource limit and reduce test costs; and, (3) the set can be scaled to have a wider coverage, if the resource limit is expanded. We propose to use the LHS (Latin Hypercube Sampling)~\cite{lhs} method to solve the sampling subproblem, as it meets all the three conditions (detailed in $\S$\ref{sec:lhsrrs}).

There also exists the second subproblem, which is to maximize the performance metric based on the given number of samples. It is required that the output configuration setting must improve the system performance than a given configuration setting, which can be the default one or one manually tuned by users. To optimize the output of a function/system, two general methods exist, i.e., model-based and search-based. Whichever method is used, the optimization must satisfy the following conditions: (1) it can find an answer even with a limited set of samples; (2) it can find a better answer if a larger set of samples is provided; and, (3) it will not be stuck in local sub-optimal areas and has the possibility to find the global optimum, given enough resources. As model-based methods generally require a large sample set, we consider search-based methods. Thus, we propose to use, along with LHS, the recursive random search (RRS) algorithm~\cite{rrs} that satisfies all the three conditions (detailed in $\S$\ref{sec:lhsrrs}).

\subsection{A Flexible Architecture}%
\label{sec:arch}

\begin{figure}[!t]
      \centering
      \includegraphics[width=0.5\textwidth]{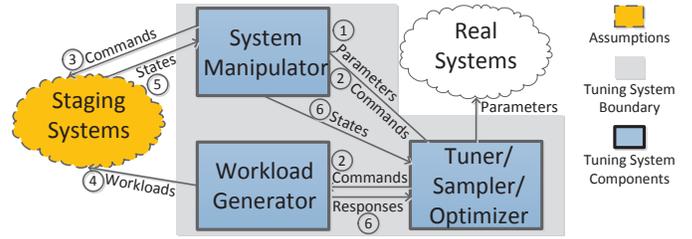}
      \caption{Architecture: automatic configuration tuning for general systems. (Best view in color)}\vspace{-12pt}
      \label{fig:act4gen} 
\end{figure}

The flexible architecture is depicted in Figure~\ref{fig:act4gen}. It mainly consists of three components, i.e., a tuner, a system manipulator, and a workload generator. Abiding by the ACTS problem definition, the tuner accepts the resource limit (typically the number of allowed tests) from the user. It extracts the configuration parameter set and their ranges from the SUT. The tuner allows different sampling and optimization methods to be used, because the SUT, the deployment environment and the workload are decoupled from the tuning process by the other two components. The workload generator allows the easy integration of various workloads for tuning, thus satisfying the workload scalability. The system manipulator can easily integrate with different SUTs in different deployment environments.

\begin{figure*}[!b]
      \centering
      \includegraphics[width=\textwidth]{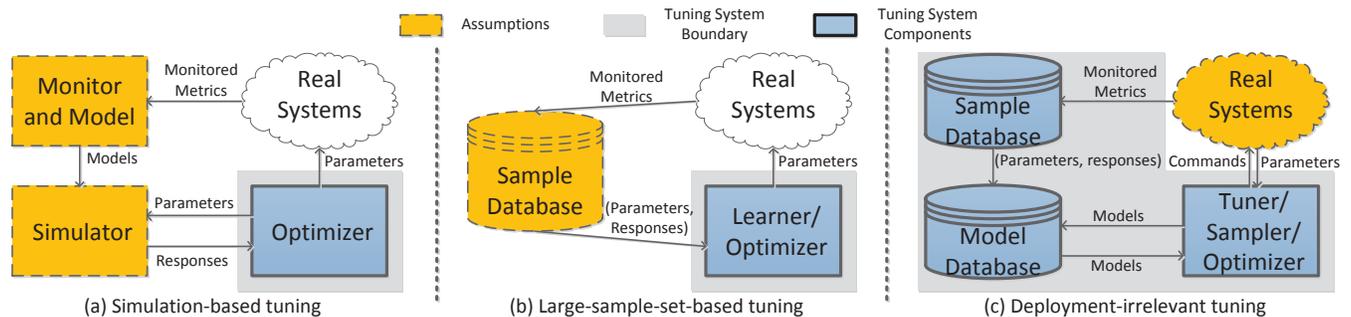}
      \caption{Common assumptions and architectures for configuration tuning in related works. (Best view in color)}
      \label{fig:otherArchs} 
\end{figure*}

Existing ACT solutions cannot solve the ACTS problem partially because their architecture designs are based on assumptions violating the scalability requirements. We can group the architectures of ACT solutions into three categories, i.e., the simulation-based architecture~\cite{paraTuning}, the large-sample-set-based architecture~\cite{aloja} and the deployment-irrelevant architecture~\cite{ottertune} that reuses samples collected from different system deployments. These architectures are illustrated in Figure~\ref{fig:otherArchs}. Explained in Section~\ref{sec:challenges}, the new challenges invalidate the assumptions that underlie these architectures.

Compared to the architectures in Figure~\ref{fig:otherArchs}, three major differences exist for the flexible architecture. First, the SUT, workload and deployment scalability is considered in the design of the workload generator and the system manipulator, with the tuner controlling these two components. Second, the tuner has not reused samples collected from other system deployments. As explained in Section~\ref{sec:challenges}, performance models are deployment-related, thus samples for other deployments cannot be reused.

Third, the tuning tests are run in a staging environment, instead of real systems or simulators. The staging environment is a mirror of the production environment, having the same actual deployment settings (e.g. hardware, clustering, software, etc.)~\cite{stagingBig,stagingWp,stagingIbm}. Using live data, it is mainly for a final test of the system before production~\cite{stagingAzure,stagingDltj}. And, implementing the real application workload in the workload generator is possible for the system in the staging environment, e.g., by log replay~\cite{replayVM,replayMr,replayCloud}. Our architecture exploits the staging environment such that samples can be collected without affecting applications on the real system deployment.
\subsection{Subproblem Solutions: LHS + RRS}%
\label{sec:lhsrrs}

Following the analysis for solving the two subproblems ($\S$~\ref{sec:design}), we adopt the LHS (Latin Hypercube Sampling)~\cite{lhs} method and the recursive random search (RRS) algorithm~\cite{rrs}.

LHS is a classic method for experimental design. Assuming that we need to collect $m$ samples. LHS divides the range of each parameter into $m$ intervals. It combines one interval of each parameter to form a subspace, in which LHS randomly chooses a sample. Repeating this process for $m$ times, $m$ samples get chosen. It is required that every interval of each parameter is used exactly once in the process.

LHS is a scalable sampling method. First, it has a wide coverage over the high-dimensional space because it considers every interval of each parameter. Second, it can sample by setting $m$ equal to the sample set constraint. Third,  if $m$ is increased, it can be scaled to have a wider coverage, because the sampling process of LHS is based on $m$.

The RRS algorithm has the exploitation and exploration structure commonly seen in search-based algorithms. In the exploration stage, RRS searches in a sample set that is taken from the whole parameter space and finds a promising sample that has the best performance. Then, it starts an exploitation stage by searching around the promising sample in the local parameter subspace. The exploitation stage is for locally searching the best point. When no improvement is made in the exploitation stage, RRS reenters the exploration stage to search globally to avoid local suboptimal results.

RRS algorithm is a scalable optimization algorithm. First, with the sample set size constraint, we can actually tune the optimization problem into one for finding a configuration setting \textbf{better} than a known setting. As a search-based method, RRS works for a sample set of any size. Second, RRS will find a better answer if a larger set of samples is provided, as it can search locally around the best sample. Third, RRS will not be stuck in local sub-optimal areas, because it has the exploration stage.

\section{How ACTS Benefits Users}%
\label{sec:benefits}

We have implemented LHS and RRS with the flexible ACTS architecture, as well as trying other sampling and optimization algorithms. We apply the ACTS implementation to MySQL and Tomcat to demonstrate how ACTS can benefit users.

ACTS can bring about the benefits of manual configuration tuning by improving system performance and increasing system utilization. Besides, due to its objectivity, ACTS also brings about the extra benefits such as enabling fairer system comparisons and identifying system bottlenecks.

\subsection{Improving System Performance:\\{\it 11 Times Better}}
\label{sec:11times}

Configuration tuning can improve the system performance, thus manual configuration tuning before using a system is in fact a common practice. General rules of "best practices" can be found on the Web for many popular systems, but they do not always provide the best results in many cases. Besides, some rules are difficult for common users to follow. As a result, although manual configuration tuning can improve system performance, users cannot always tune a system to the system's best potential.

ACTS only changes the configuration settings of a system, but the possible performance gain can be as much as \textbf{11 times}. In the example of MySQL, the best configuration setting suggested by ACTS can reach a throughput of 118184 ops/sec, while that for the default setting is only 9815 ops/sec. In comparison, many systems implementing new designs can only improve the system performance by \emph{a limited percentage or multiple times.} That is, an easy change of the configuration settings can benefit the user much more than laboriously implementing new designs. Moreover, as many workloads are repetitive and recurring~\cite{repeatWork2,repeatWork1}, this performance gain can actually be highly significant to users.
\begin{table}[b]
  \centering
  \small
  \renewcommand\arraystretch{1.1}
  \caption{ACTS improving performances of a fully-utilized Tomcat server.}\vspace{-3pt}
  \label{tbl:tomcatOnArm}%
  \begin{tabular}{lllc}
\toprule[1pt]
 { \textbf{Metrics}} & { \textbf{Default}} & { \textbf{BestConfig}} & \textbf{\small Improvement}\\
  \midrule[0.8pt]
    Txns/seconds& 978 & 1018 & $\mathbf{4.07\%\uparrow}$\\
      \midrule[0.2pt]
    Hits/seconds& 3235 & 3620 &  $\mathbf{11.91\%\uparrow}$\\
      \midrule[0.2pt]
    Passed Txns& 3184598& 3381644&  $\mathbf{6.19\%\uparrow}$\\
      \midrule[0.2pt]
    Failed Txns& 165& 144&  $\mathbf{12.73\%\downarrow}$\\
      \midrule[0.2pt]
    Errors& 37& 34&  $\mathbf{8.11\%\downarrow}$\\
\bottomrule[1pt]
\end{tabular}
\end{table}
\subsection{Improving System Utilization:\\{\it Eliminating 1 from every 26}}\vspace{2pt}

ACTS can also improve system utilization by reducing the demands of virtual machines. Nowadays, it is common that many systems are deployed on the virtual machines in the cloud. Improving the throughputs of a single virtual machine can in turn reducing the number of virtual machines in need.

In a use case of Tomcat, we apply ACTS to Tomcat servers deployed on virtual machines, which run on physical machines equipped with ARM CPUs. Each virtual machine is configured to run with 8 cores, among which four are assigned to process the network communications. Under the default configuration setting, the utilizations of the four cores serving network communications are fully loaded, while the utilizations of the other four processing cores are about 80\%. By automatic configuration tuning, a better configuration setting is found to improve the performance of the deployment by 4\%, while the CPU utilizations remain the same. The performance results of the tuned and the default configuration settings are presented in Table~\ref{tbl:tomcatOnArm}. We can observe improvements on every performance metric by the tuned configuration setting. With this improvement on throughput, we can eliminate 1 virtual machine from every 26 virtual machines, if the tuned configuration setting is used instead.\vspace{1.5pt}

\subsection{Saving Labor Costs:\\{\it Machine-Days vs. Man-Months}}\vspace{1.5pt}

Configuration tuning is highly time-consuming and laborious. It requires the users: 1) to find the heuristics for tuning; 2) to manually change the system configuration settings and run workload tests; and, 3) to iteratively go through the second step many times till a satisfactory performance is obtained. Sometimes, the heuristics in the first step might misguide the users, as some heuristics are correct for one workload but not others; then, the latter two steps are in vain.

In our experience with MySQL tuning, it has once taken five junior employees about half a year to find an appropriate configuration setting for a cloud application workload. We have also exploited our ACTS system to tune the same system deployment. A better performance is achieved within two days. Automatic configuration tuning not only saves labor costs, but also shortens the tuning time from months to days. Even if system experts might tune a system much better and faster than common users, they are very expensive to hire~\cite{expensive}. In comparison, automatic configuration tuning almost involves no labor costs.

\subsection{Fairer Benchmarking and Comparison of Systems}\vspace{1.5pt}

Benchmarking is a well-established method for comparing the performance of various hardware or software systems~\cite{greybenchmark,berkeleyview}, e.g., running SPEC for hardware comparison~\cite{spec} or TPC benchmarks for database systems~\cite{tpc}.To enable an \emph{apples-to-apples} comparison between systems, it is required that the only changed factor on benchmarking is the system under test. Besides, to get a good benchmarking result, the system under test must be well tuned~\cite{fairbenchmarking,tuneb4benchmark}. Configuration tuning is part of the performance tuning process. However, the performance tuning process is highly subjective and depends heavily on the tuning experts.

ACTS enables an objective tuning process and enables fairer starting points for benchmarking. As demonstrated by the MySQL case, a simple change of parameters can lead to more than 11 times performance improvement. As many new system designs can only improve system performances by some percentage or multiple times, it is more relevant that any improvement on system be tested on an optimized system state. Without a proper configuration tuning process, the benchmarking results can be highly suspicious or misguiding. As previous configuration tuning is usually manual work, there is no way to define how a system is in a state ready for benchmarking. To tap the performance potential of a system, system users need help in configuration tuning.\vspace{1.5pt}

\subsection{Identifying System Bottlenecks}\vspace{1.5pt}

In the use case of Big Data, it is common that multiple systems are deployed simultaneously for an application. For example, we might need to deploy the Hadoop file system for using Spark, or run a workload balancing system to distribute requests to the backend database system. Among the co-deployed systems, we might need to find out which system is the bottleneck in order to improve the overall performance.

ACTS can help identify system bottlenecks by (1) tuning the subsystem to its best performance; and, (2) combine systems to tune for the best performance. Take the database system for example. Database is usually deployed along with a front-end caching and load balancing system. Once we have tuned a database system by itself and improved the performance by 63\%. Then, we apply the same workload to the tuned database system through a front-end caching and load balancing system. Even after a long time tuning, we found that the performance remaind at the untuned level for the database system co-deployed with the front-end system. By such, we located the bottleneck to be the front-end caching and load balancing system. Without automatic configuration tuning, we would not be able to make sure whether the reason is configuration setting or systems themselves.

Furthermore, by automatically tuning each system or the system combination to its best performance, we can also identify the bottleneck to be a specific system, if the system has the worst performance among all systems and system combinations; or, if the system combination has the worst performance, the bottleneck is the specific system combination. When a combination of systems has the worst performance, it indicates that the member systems are having interactions affecting the overall performance. This bottleneck identification can help users decide whether to improve the design of a specific system or to reduce the influences between systems.\vspace{3pt}

\section{Conclusion}

In this paper, we comprehensively investigate the challenges and analyze the characteristics of the ACTS problem. The solution to the ACTS problem must guarantee scalability with regard to resource limit, configuration parameter set, SUT, deployment environment and workload. We propose and implement a preliminary ACTS solution. This solution features a flexible architecture, which enables the easy integration of various SUTs, deployment environments and workloads, as well as scalable sampling methods and optimization algorithms. The scalable sampling method and optimization algorithm adopted in the preliminary solution are LHS and RRS respectively. Based on the initial experimental results, we demonstrate that ACTS can benefit users in facilitating the system usage, improving the system performance, increasing the system utilization, saving labor costs, fairer benchmarking results, system bottleneck identification, etc.

Systems are becoming more complex nowadays. We believe that ACTS will become more beneficial or even indispensable to users. As a result, we believe that future systems should be equipped with automatic configuration tuning. We have only proposed a preliminary solution to the ACTS problem to demonstrate that ACTS is solvable. Great research opportunities exist in devising better solutions to ACTS and equipping systems with ACTS.\vspace{6pt}

\section*{Acknowledgments}

We would like to thank our shepherd, Cheng Li, and the anonymous reviewers for their constructive comments and inputs to improve our paper. This work is in part supported by the National Natural Science Foundation of China (Grant No. 61303054), the State Key Development Program for Basic Research of China (Grant No. 2014CB340402) and gifts from Huawei.

\balance

\bibliographystyle{ACM-Reference-Format}
\bibliography{ref}


\end{document}